# Magnetic cooling at Risø DTU

K.K. Nielsen[a,b], R. Bjørk[b], J.B. Jensen[b], C.R.H Bahl[b], N. Pryds[b], A. Smith[b], A. Nordentoft[b], J. Hattel[a]

[a] Technical University of Denmark, Department for Mechanical Engineering, Produktionstorvet, building 425 room 024, 2800 Kgs. Lyngby, Denmark, kaspar.kirstein.nielsen@risoe.dk

[b] Department for Fuel Cells and Solid State Chemistry, Risø National Laboratory for Sustainable Energy, Technical University of Denmark - DTU, building 232, Frederiksborgvej 399, 4000 Roskilde

## ABSTRACT

Magnetic refrigeration at room temperature is of great interest due to a long-term goal of making refrigeration more energy-efficient, less noisy and free of any environmentally hostile materials.

A refrigerator utilizing an active magnetic regenerator (AMR) is based on the magnetocaloric effect, which manifests itself as a temperature change in magnetic materials when subjected to a varying magnetic field.

In this work we present the current state of magnetic refrigeration research at Risø DTU with emphasis on the numerical modeling of an existing AMR test machine. A 2D numerical heat-transfer and fluid-flow model that represents the experimental setup is presented. Experimental data of both no-heat load and heat load situations are compared to the model. Moreover, results from the numerical modeling of the permanent magnet design used in the system are presented.

## 1. INTRODUCTION

The magnetocaloric effect (MCE) was discovered by E. Warburg in 1881. Warburg found that iron got heated up when placed in a magnetic field and when the magnetic field was removed the iron sample cooled down (Warburg 1881). The basic principle of the MCE is that the ordering of the magnetic moments is increased when an external magnetic field is applied to a magnetic material. This means that the spin-entropy decreases. The process is virtually adiabatic if the field is applied rapidly. This means that the total entropy of the system must remain constant and thus the lattice and electron entropies must increase, which is equivalent to an increase in temperature. The process is reversible (for some materials) and thus the opposite will take place when the field is removed again (i.e. the ordering of the magnetic moments decrease and the temperature thus decreases). The MCE is strongest at the phase-transition between the ferromagnetic and the paramagnetic phases. This phase transition takes place at the Curie temperature $T_C$, which can vary significantly depending on the material. In the past materials have been used mainly for cryogenic applications, but some 30 years ago research into the MCE at room temperature was commenced (Brown 1976) .

The MCE yields, for the benchmark magnetocaloric material (MCM) gadolinium (Gd), an adiabatic temperature change of about 3.6 K at room temperature for a 1 tesla (T) magnetic flux density. This rather low temperature change is obviously too small for direct usage in a cooling device. However, if the material is used in an AMR it is possible to achieve, due to regeneration, a higher temperature difference (Brown 1976). In his experiments Brown reached a temperature span of 46 K using Gd with the hot end at 319 K using a 7 T magnetic flux density from a super conducting magnet. The MCE of Gd is proportional to the magnetic flux density to the power of 0.7 (Pecharsky and Gschneidner 2006). Today's state-of-the-art permanent magnets yield a magnetic flux density of about 1.5 T (Tura and Rowe 2007). Therefore it is crucial to develop a high-performing and efficient AMR.

This work is primarily concerned with developing a model describing an existing AMR test machine based on parallel plates, and using a permanent magnet based on the Halbach design yielding around 1.1 T (Halbach 1980) . In Section 2 the experimental test machine is described. In Section 3 the corresponding numerical model is presented. In Section 4 results from the test machine and the model are compared both including no-load and load-situations. In Section 5 the results are discussed and the work is concluded with some future aspects briefly discussed.

## 2. EXPERIMENTAL SETUP

Figure 1 shows photos of the test machine, which consists of a regenerator core in the middle of a plastic tube with outer diameter 40 mm and inner diameter 34 mm. The regenerator core is built up of 13 plates of 99.9 % pure Gd (obtained from China Rare Metal Materials Co). The plates with dimensions 40x0.9x25 mm have a total mass of 92 g. At both ends of the Gd plates (in the flow direction) 20 mm long plastic flow guides are placed to ensure a fully developed laminar flow across the plates. The plates and flow guides are fixed by precision machined grooves and are stacked with a spacing of 0.8 mm, which is then the height of the fluid channel.

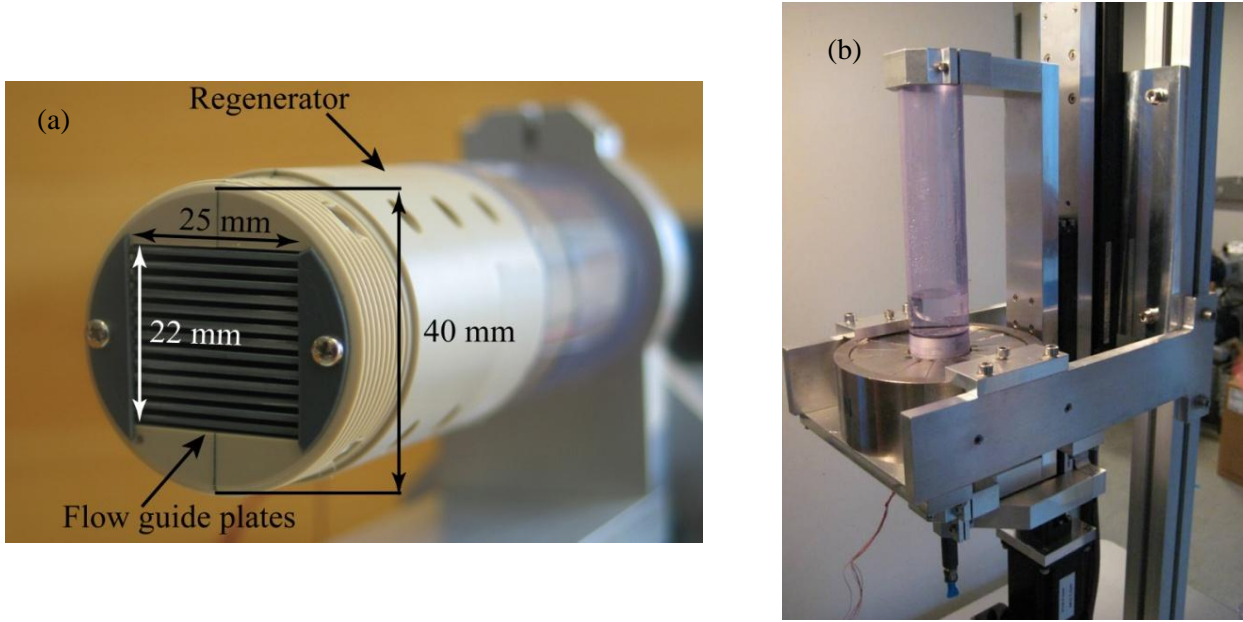

*Figure 1: Figure (a) shows a close-up of the experimental AMR test machine where the 13 parallel channels can be seen as well as the plastic tube. Figure (b) is a picture of the machine in its operational environment. The permanent Halbach magnet can be seen with the plastic tube including the regenerator core penetrating it.*

The heat transfer fluid is moved by a piston. The regenerator block and its parent plastic tube are suspended vertically in a mounting as shown in Figure 1b and can be moved in and out of the field of the permanent Halbach magnet using stepper motors. This magnet has a maximum magnetic flux density of 1.1 T.

One of the most important results of the experiment – as well as in the model – is to be able to measure the temperature gradient across the regenerator core. This is done via five type E thermo-couples placed equidistantly in the center flow channel as sketched in Figure 2a.

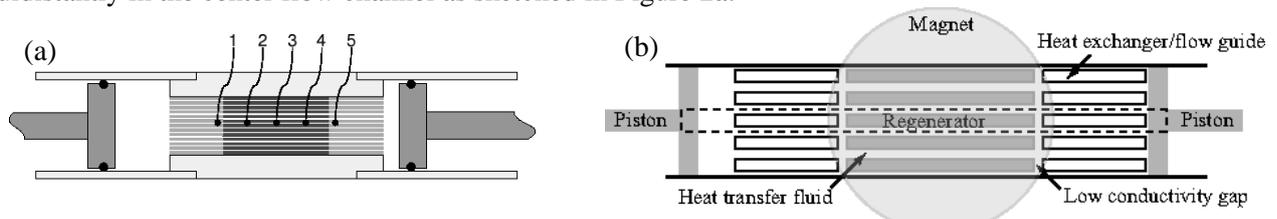

*Figure 2: Drawing (a) is a schematic of the regenerator pictured in Figure 1a. The locations of the five thermo-couples are indicated with their appropriate numbers. Thermo-couples 1and 5 are placed at the cold and hot ends respectively. Figure (b) shows how the numerical model represents the full geometry of the AMR. The model breaks the geometry down into a single replicating cell consisting of one half of a complete flow channel (indicated with a dashed line in the figure and magnified in Figure 3).*

The system evolves transiently through a number of AMR cycles until cyclic steady-state has been reached. Each cycle consists of four different steps, which have four different characteristic times $\tau_1$, $\tau_2$, $\tau_3$ and $\tau_4$. The cycle is symmetric meaning that $\tau_1 = \tau_3$ and $\tau_2 = \tau_4$. In the first step the magnetic field is applied thus increasing the temperature of the MCM and at this stage the fluid is stationary. In the second step, the pistons move the fluid for $\tau_2$ seconds towards the hot end of the regenerator to reject heat. At the third step the magnetic field is switched off and thus the temperature in the MCM decreases and again at this stage the fluid is stationary. Finally, the piston pushes the fluid towards the cold end for $\tau_4$ seconds. The total cycle-time is $\tau_{total} = 2(\tau_1 + \tau_2)$. In this way the MCM is used as the active material in a regenerator and a temperature gradient is built up. The magnitude of this gradient depends mainly on the geometry, material and operational properties, i.e. the piston stroke length, $\tau_1$ and $\tau_2$, the height of the fluid channel, the MCM, and how strong the magnetic field is. It is therefore quite a challenge to predict the behavior of a certain system for different process parameters.

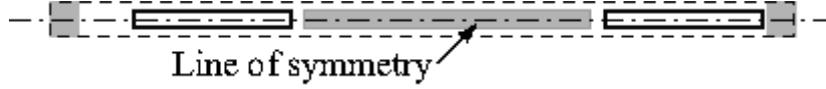

*Figure 3: A close-up of the line of symmetry from the replicating cell marked with a dashed line in Figure 2b.*

The geometrical simplicity of such an experimental setup makes it ideal for studies of parallel plate regenerators, facilitating direct comparison to the numerical model. Validating the model against the experiment is crucial since a high-quality model can predict the performance of configurations otherwise not thought of and span a much larger parameter-space than possible with the experiment.

## 3. NUMERICAL MODELING

### 3.1 Thermal model of the regenerator

The numerical model is "2.5-dimensional" as illustrated geometrically in Figure 4 and Figure 5. For technical reasons the heat transfer fluid is chosen to be stationary and the solid domains are moved relative to this. Thus, the piston movement is modeled as a coordinate transformation of the solid domains with a suitable convective term in the thermal equation for the fluid. The spatial discretization is the classical $2^{nd}$ order finite difference scheme with a equidistant grid where $\Delta x = 1$ mm and $\Delta y = 0.05$ mm, and the temporal integration is done using an Alternate Direction Implicit (ADI) solver with a timestep chosen to be 0.001 second. Since the system includes moving boundaries it is extremely important to make sure that there is energy conservation. Therefore the finite difference (FD) formulation is preferred and validation-tests show that the energy-conservation is virtually the precision of the computer. The computational time on a 2.0 GHz Intel Core 2 Duo CPU is roughly 0.7 CPU-seconds pr physical second in the model.

Due to symmetry considerations only half a replicating cell is modeled (as indicated in Figure 3). This is a good assumption at least for the central channels and plates (which have virtually no loss through the top and bottom of the regenerator).

Figure 4a and Figure 5 show a schematic of the boundary conditions of the model in the (x,y)-plane and (x,z)-plane respectively. The various thermal resistances are labeled with their respective names.

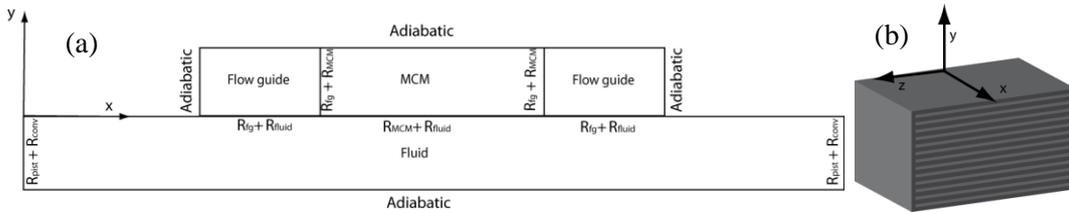

*Figure 4 : Figure (a) shows a schematic of the modeled domain in the (x,y)-plane, i.e. half a replicating cell with the boundaries being either adiabatic (symmetry boundaries) or coupled via thermal resistances to the ambient. The x-direction is the direction of the flow and the y-direction is orthogonal to the plates (labeled MCM). The left end is defined as the cold end and the right end as the hot end. Figure (b) shows a 3D sketch of the regenerator block with the coordinate system visualized.*

The governing equations for the thermal system are

$$\frac{\partial T_{fl}}{\partial t} = \frac{k_{fl}}{\rho_{fl} c_{p,fl}} \nabla^2 T_{fl} - (\boldsymbol{u} \cdot \nabla) T_{fl} \tag{1}$$

$$\frac{\partial T_s}{\partial t} = \frac{k_s}{\rho_s c_{p,s}} \nabla^2 T_s \tag{2}$$

where the temperatures of the fluid and solid domains are denoted by $T_{fl}$ and $T_s$ respectively. For simplicity all the solid domains are labeled with an s, although they have different physical properties. The thermal properties, i.e. the thermal conductivities $k_{fl}$ and $k_s$, the mass densities $\rho_{fl}$ and $\rho_s$ and the heat capacities $c_{p,fl}$ and $c_{p,s}$ are all assumed constant except the heat capacity of Gd, which varies as function of both temperature and magnetic field (see Figure 6). The material properties used are given in Table 1.

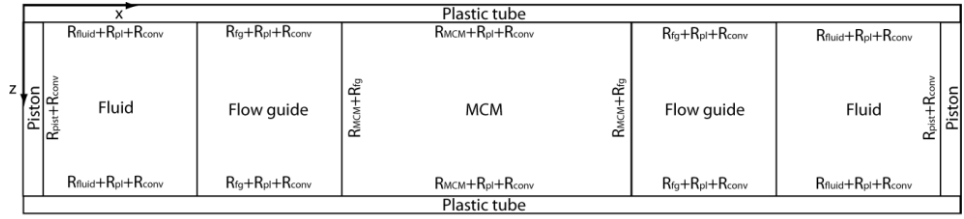

*Figure 5 : The model in the (x, z)-plane. The z-direction is only resolved by one grid cell meaning that the model is effectively 2.5-dimensional with the x- and y-dimensions being the two regular dimensions and the finite extension of the z-direction as the half dimension (and most importantly including losses via boundary conditions).*

The velocity field in the fluid is denoted by $\mathbf{u} = (u, v)$ and is prescribed by the analytical expression for a parallel-plate laminar flow with piston velocity $u_p$, see e.g. (T. F. Petersen 2007):

$$u = \frac{H_{fl}^2}{2\mu} \frac{\partial p}{\partial x} \left(1 - \frac{y^2}{H_{fl}^2}\right) + u_p \quad (3)$$

$$v = 0 \quad (4)$$

$$\frac{\partial p}{\partial x} = \frac{96}{Re} \rho_{fl} \frac{1}{4H_{fl}} \frac{u_p^2}{2} \quad (5)$$

The Reynolds' number $Re = u_p 4H_{fl}\rho_{fl}/\mu$, $\rho_{fl}$ is the mass density of the fluid, $H_{fl}$ is half the height of the fluid channel, $\mu$ is the viscosity of the fluid and $y$ is the vertical coordinate, i.e. orthogonal to the flow direction.

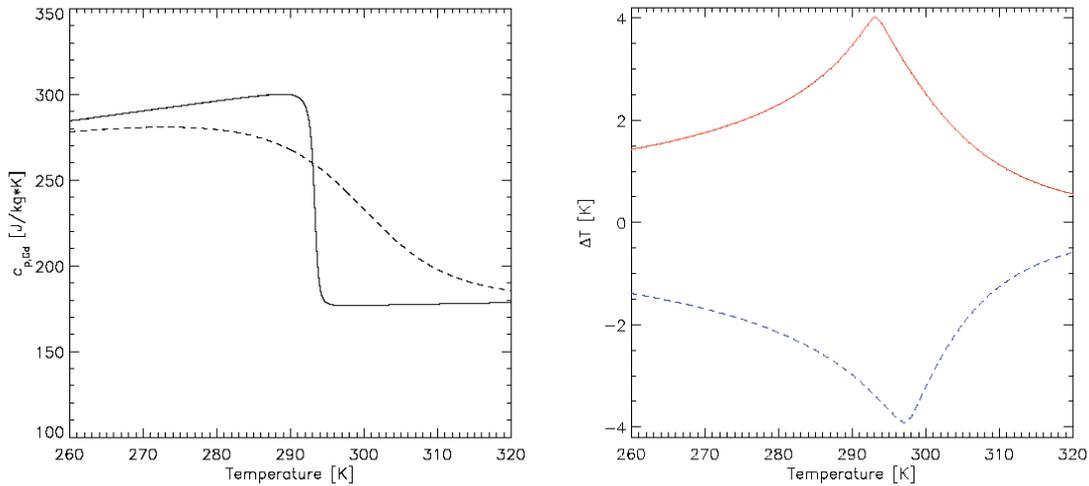

*Figure 6 : Left: $c_p$ for Gd as function of temperature in zero field (solid line) and in a 1 T field (dashed line). The change around 293 K is rather significant and is actually the definition of the Curie temperature. Right: The adiabatic temperature change of Gd around room-temperature in a 1 T field. The red/solid line is the temperature increase when the field is applied and the blue/dashed line is the corresponding curve for when the field is removed. The data are calculated from the mean field model of Gd compiled in e.g. (Petersen, et al. 2008).*

The internal boundaries between the fluid domain and the solid domains are implemented through thermal resistances in Fourier's law of thermal conduction:

$$q_{bd} = -\frac{T_1 - T_2}{R_1 + R_2}. \quad (6)$$

Here the flux across the boundary between two domains (e.g. fluid and MCM) is denoted by $q_{bd}$, the temperature of the boundary cells in the two adjacent domains are $T_1$ and $T_2$ and their corresponding thermal resistances are $R_1$ and $R_2$ respectively. The thermal resistance is simply given by the distance from the grid cell's centre to the boundary face divided by the thermal conductivity of the material multiplied by the area of the face boundary.

*Table 1 : Material properties used in the model obtained from (Petersen, et al. 2008) and (Holman 1987).*

| Material | $k$ [W/m·K] | $\rho$ [kg/m$^3$] | $c_p$ [J/kg·K] | $\mu$ [kg/m·s] |
|---|---|---|---|---|
| Water/ethanol mixture | 0.52 | 981 | 4330 | $8.91 \cdot 10^{-4}$ |
| Plastic | 0.2 | 1200 | 840 | n/a |
| Gd | 10.5 | 7900 | 170-300 | n/a |

The outer boundaries are either adiabatic, if they are symmetry boundaries, or they simulate heat loss in the z-direction, which is not directly resolved (hence this is what we call a 2.5-dimensional model). These losses are calculated via thermal resistances and they contain the thicknesses and thermal conductivities of the particular domain (fluid or solid) and the insulating material surrounding the entire system. On the outer part of the insulating material there is assumed to be natural convection modeled via the parameter $h_{conv}$, which has a value in the range $5 - 20$ W/Km$^2$ and corresponds to free convection of air on a plate (Holman 1987).

### 3.2 The permanent magnet

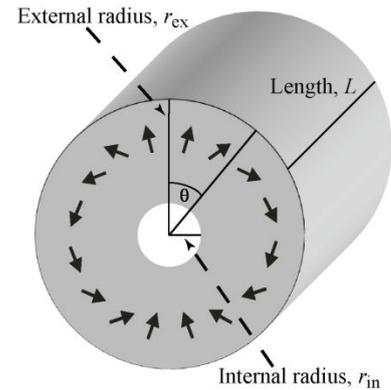

The magnetic field that generates the MCE can be produced by an electromagnet or a permanent magnet assembly. For this machine we have chosen the latter as this requires no external power source to produce a strong magnetic field. The requirement of the permanent magnet assembly is that it must produce a strong homogenous magnetic field in a confined region of space and a very weak field elsewhere. The design known as a Halbach cylinder (Mallinson 1973), (Halbach 1980) fulfills these requirements and has therefore been chosen for the test machine. An ideal Halbach cylinder consists of a permanent magnetic material with a bore along the cylinder symmetry axis. The magnetic material is magnetized such that the direction of magnetization varies as shown in Figure 7. This produces a strong homogeneous field in the cylinder bore. In the case of an infinitely long cylinder the flux density in the bore is given by $B = B_r \ln\left(\frac{r_{ex}}{r_{in}}\right)$. An ideal Halbach cylinder is not physically realizable, as it is both necessary to make the Halbach cylinder of a finite length and to divide the continuously magnetized cylinder into parts consisting of permanent magnets each with their own directions of magnetization.

*Figure 7: A drawing of a Halbach cylinder showing the internal radius, $r_{in}$, external radius, $r_{ex}$, and length, L. Also shown are arrows in the direction of the remanent magnetization of the magnetic material. This varies as $2\theta$. The figure is from (Bjørk, et al. 2008)*

Based on the design of the regenerator the Halbach cylinder for the test machine consists of 16 blocks of permanent magnets and with dimensions $r_{in} = 2.1$ cm, $r_{ex} = 6$ cm, and $L = 5$ cm.

To investigate the magnetic field produced by this Halbach cylinder we have performed numerical simulations using the commercially available finite element multiphysics program, *Comsol Multiphysics* (Comsol 2005), see also (Bjørk, et al. 2008) for details.

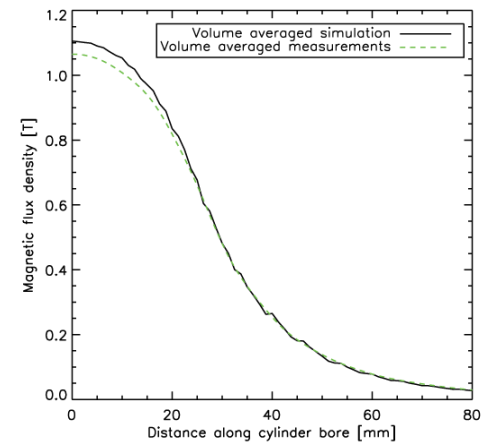

*Figure 8: Flux density for the simulated and the physical Halbach cylinder for the test machine. There is good agreement between data.*

As well as modeling the magnet assembly we have also performed measurements of the flux density of the physical magnet assembly, seen in Figure 1. In Figure 8 the average flux density of the magnetic field as a function of distance from the center of the Halbach cylinder for both simulation and measurement is shown. As can be seen from the figure the numerical simulation and the experimental measurements agree, and show that a high flux density is produced in the center of the cylinder bore.

## 4. RESULTS

The experimental and modeling results are divided in two parts. First a sensitivity analysis of how far the regenerator is taken out of the Halbach's magnetic field is addressed under no-load conditions. Secondly a load-situation is investigated.

### 4.1 Sensitivity to the magnetic field

Since the magnetic field of the Hallbach magnet strays outside of the central bore in the cylinder (see Figure 8), the distance which the regenerator block is moved away from the centre of the Halbach must have some influence on the performance of the regenerator. The experiments were adjusted to move the regenerator out of the magnetic field with a distance varying from 30 mm to 150 mm (see Figure 9). The operating conditions were the same for each experiment, which was allowed to reach steady-state in each case (see Table 2). The model was set with the same parameters and the varying magnetic field was implemented via a volumetric source term in the heat equation for the MCM:

$$\frac{dQ_{MCM}}{dt} = -\rho_{Gd} T_{Gd} \frac{\partial \sigma}{\partial T} \frac{dB}{dt}. \qquad (7)$$

This is obtained from the mean field theory of Gd, see e.g. (Petersen, et al. 2008). The change with respect to temperature of the magnetization is denoted by $\partial\sigma/\partial T$ and the magnetic flux density is denoted by $B$. The magnetic field only varies in the x-direction in the regenerator. The crucial term in this formulation is the time variation of the magnetic field. This is implemented simply using the finite extent of the regenerator block and the velocity of which the regenerator is moved in and out of field.

As seen in Figure 9 there is one series of experimental data and two model series. The data sets show the no-load steady-state temperature span between thermo-couples one and five as function of how far the regenerator is taken out of the magnetic field. It is seen from the experimental data that at distances above 70 mm the temperature span does not increase anymore; hence, the full yield of the magnet is utilized.

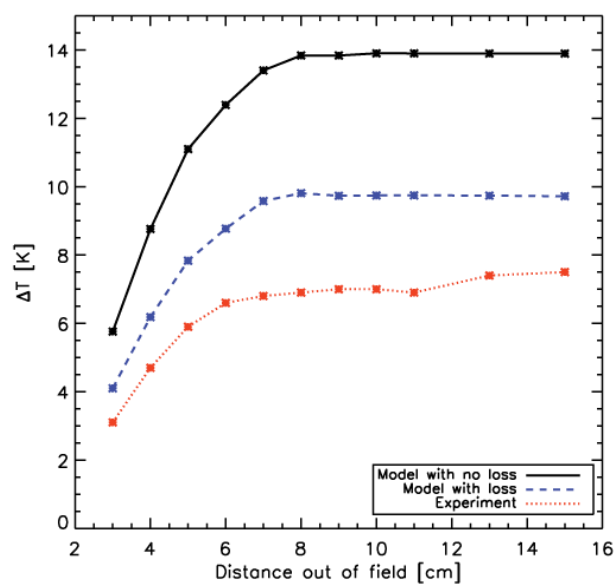

*Figure 9: The figure shows how the steady-state and no-load temperature span behaves when the regenerator is not taken completely out of the magnetic field (the red/dotted line). Each asterisk in the graph represents a data point. Also included are the results of two slightly different numerical simulations; one without losses (the black/solid line) and one with ideal losses (blue/dashed line). The tendencies are clearly the same on all three graphs. The absolute values of the temperature spans differ somewhat, however, including losses is seen to improve the correspondence between experiment and model significantly.*

The model simulations were done for two cases: One with no loss to the surroundings, i.e. perfect thermal insulation, and one with realistic losses via the boundary conditions described in Section 3.1. The tendencies of all three data sets are virtually the same, which clearly shows that the numerical model catches many of the aspects of the magnetic regeneration. It is not surprising that the ideal adiabatic model overestimates the temperature span somewhat as significant losses to the ambient are expected in the test device. When the losses are included, however, the model comes much closer at the experimental values still showing the exact same tendency.

*Table 2 : The operational properties of the two experiment series.*

| Experiment | Effective piston stroke length (% of plate length) | $\tau_1$(s) | $\tau_2$(s) |
|---|---|---|---|
| Magnetic field variation | 40 % | 3.0 | 2.9 |
| Heat load experiment | 53 % | 1.5 | 2.9 |

### 4.2 Load experiment

The piston at the cold end has been equipped with a copper plate connected to a power supply which makes it possible to apply a heat load through ohmic dissipation to the water. An experiment was run with the parameters given in Table 2 and heat loads from 0 to 1.6 W. The model was set with the same parameters and a spatially constant magnetic flux density of 1 T. Figure 10 shows both an example of the transient evolution of a specific heat load experiment (left-hand) and the results of the heat-load series (right-hand).

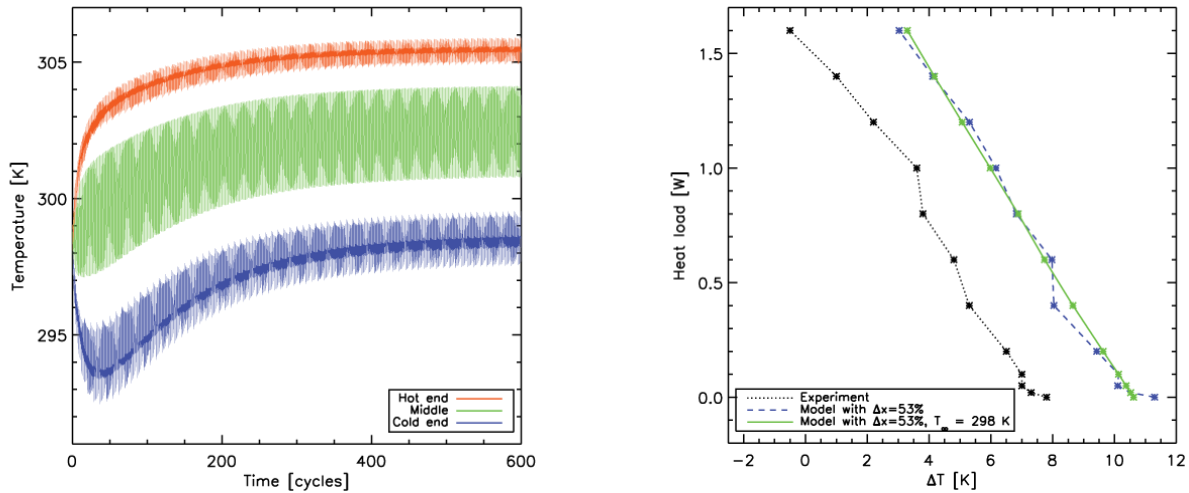

*Figure 10: Left: The transient evolution of the cold, middle and hot parts of the regenerator (simulated). The particular example is for a piston stroke ($\Delta x$) of 53% with a load of 0.8 W. Right: A load-experiment and the corresponding model results. The model assumed $h_{conv} = 20\ W/m^2K$. Note that there are two model-series in the right graph. The green/solid line data set was performed with a constant ambient temperature whereas the blue/dashed line data set corresponds directly to the circumstances during the experimental data acquisition (black line/dotted).*

The experimental series was performed over a period of two days since it takes around an hour to reach steady-state for each configuration. Therefore the ambient temperature $T_\infty$ varied slightly (from 296-299 K). This is possible to adjust in the model as well, and therefore the two data sets are directly comparable. The model and the experimental data are very similar in behavior, though the model over-estimates the temperature span. Generally the temperature span decreases linearly with the increasing cooling capacity as one would expect. There are, however, minor fluctuations in the linearity. If the experimental data are considered isolated, the small variations may be regarded as experimental noise. However, when compared to the model data, virtually the same variations are seen. To investigate this, a model-series was performed with the ambient temperature set to the constant value 298 K. This is seen as the green/solid line in the right graph of Figure 10. Thus, the variations away from the linearly decreasing cooling capacity are interpreted as a result of the fluctuations in the ambient temperature. The slopes of each of the three graphs were found by linear regression. The values are all -0.2 $\pm 0.01$ W/K.

## 5. DISCUSSION, CONCLUSIONS AND OUTLOOK

### 5.1 Discussion

The numerical model has been successfully validated against real experiments in different situations including no-load and load-experiments, varying the magnetic field and some of the operational parameters, namely piston stroke length, $\tau_1$ and $\tau_2$. The discrepancies between the model and the experiment seen in Figure 9 and Figure 10 are, however, something that should be considered and the model should be improved to minimize these. We have used an ideal model for the behavior of Gd in terms of $c_p$ and $\Delta T_{ad}$. We have independently measured the actual adiabatic temperature change of the Gd used in the test machine and it has turned out that due to impurities the actual adiabatic temperature change is roughly 20 % lower than in the ideal mean field model used in the numerical model. We have chosen not to include this in the present work since we have not yet performed enough measurements of the utilized Gd in order to cover the range in magnetic fields and temperature span needed.

A result of this work is that the model is directly capable of catching the effect of the ambient temperature on the system. This may have been interpreted as an experimental feature (e.g. noise) if the model had not caught it and if not the constant-ambient temperature modeling had resulted in the completely straight line seen in the right part of Figure 10.

## 5.2 Conclusions and outlook

The experimental AMR at Risø DTU has been demonstrated to be quite versatile in terms of operational parameters and various aspects of the cooling capacity. The corresponding numerical model is to a large extent successful in predicting the behavior of the system. Many interesting aspects still need to be investigated though. They include obtaining more reliable and realistic data of the Gd we actually use in our test machine, testing other potential MCM materials and changing the thickness of the plates and the fluid channels as well as the operating parameters. Having a powerful numerical model that predicts the behavior seen experimentally is crucial for the further development of a new AMR with significantly improved performance. The fact that there is a very strong correspondence between the experimental and modeling results in both series presented in Section 4 strongly indicates that the model indeed captures the general behavior of the parallel-plate AMR system.

## ACKNOWLEDGEMENTS

The authors thank Mr. Jørgen Geyti for his technical assistance. Furthermore the authors would like to acknowledge the support of the Programme Commission on Energy and Environment (EnMi) (Contract no. 2104-06-0032) which is part of the Danish Council for Strategic Research.